\documentclass[letterpaper, 10 pt, conference]{ieeeconf}
\IEEEoverridecommandlockouts                              
\usepackage[utf8]{inputenc}
\usepackage[T1]{fontenc}
\usepackage{graphicx, xcolor}
\usepackage{mathtools}
\usepackage{times}
\usepackage{hyperref}
\usepackage{fancyhdr,graphicx,amsmath,amssymb}
\usepackage[ruled,vlined]{algorithm2e}
\include{pythonlisting}

\title{\LARGE \bf
The Bloom Tree
}

\author{Lum Ramabaja \\ \href{mailto:lum@bloomlab.io}{lum@bloomlab.io} 
   \and Arber Avdullahu \\ \href{mailto:lum@bloomlab.io}{arber@bloomlab.io} 
}

\begin{document}

\maketitle

\begin{abstract}

We introduce a data structure that allows for efficient (probabilistic)  presence proofs and non-probabilistic absence proofs in a bandwidth efficient and secure way. The Bloom tree combines the idea of Bloom filters with that of Merkle trees. Bloom filters are used to verify the presence, or absence of elements in a set. In the case of the Bloom tree, we are interested to verify and transmit the presence, or absence of an element in a secure and bandwidth efficient way to another party. Instead of sending the whole Bloom filter to check for the presence, or absence of an element, the Bloom tree achieves efficient verification by using a compact Merkle multiproof.

\end{abstract}

\section{Introduction}
\label{s:intro}

\subsection{Bloom filters}
\label{sb:Bloom_filters}

A Bloom filter is a space-efficient probabilistic data structure that allows to verify if an element is \textit{not} in a set. In other words, a Bloom filter can either tell us that an element \textit{might} be in a set, or that an element definitely is \textit{not} in a set. Bloom filters, and variants of the Bloom filter have found a wide range of applications - They have been extensively used in blockchains \cite{Ma2019BlockchainFilter}, for set reconciliation problems \cite{Guo2013SetFilters}, for memory-efficient genome assembly \cite{Jackman2017ABySSFilter}, and more. The main reason why Bloom filters are used in so many domains are their very low space complexity.

\begin{figure}[h]
\centering\includegraphics[width=0.7\linewidth]{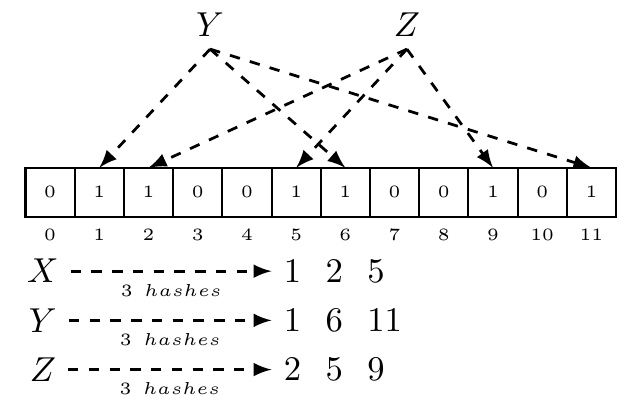}
\caption{Inserting multiple elements into a Bloom filter. ''\textit{X}'' was never inserted, yet it appears as if it has, i.e. a false positive has occurred.}
\label{fig:bf}
\end{figure}

To populate a Bloom filter, we first initiate a zero bit array. Whenever we want to insert an element to the Bloom filter, we hash the element $k$ times (as in figure \ref{fig:bf}), go to the indices to which the hashes point, and turn those values to one. When checking for the presence of an element in the Bloom filter, we simply check if all the given indices have a one. If one of the indices is still a zero, we know that the element was never inserted into the Bloom filter. The false positive nature of the Bloom filter comes from the fact that the values of some indices for a given element might already be turned to one, even though we never inserted the element. An observer looking at figure \ref{fig:bf} for example might think that element ''X'' was inserted to the Bloom filter, even though it was not.

The false positive rate of a Bloom filter can be controlled by modifying three variables:
\begin{enumerate}
    \item The Bloom filter size $m$.
    \item The number of elements inserted into a Bloom filter $n$.
    \item The number of hash functions used per element $k$.
\end{enumerate}{}
The formula of which can be written as:

\begin{equation}
\label{eq:bf_fpr}
\left(1-\left(1-\frac{1}{m}\right)^{kn}\right)^k
\end{equation}

\subsection{Merkle trees}

\begin{figure}[h]
\centering\includegraphics[width=0.7\linewidth]{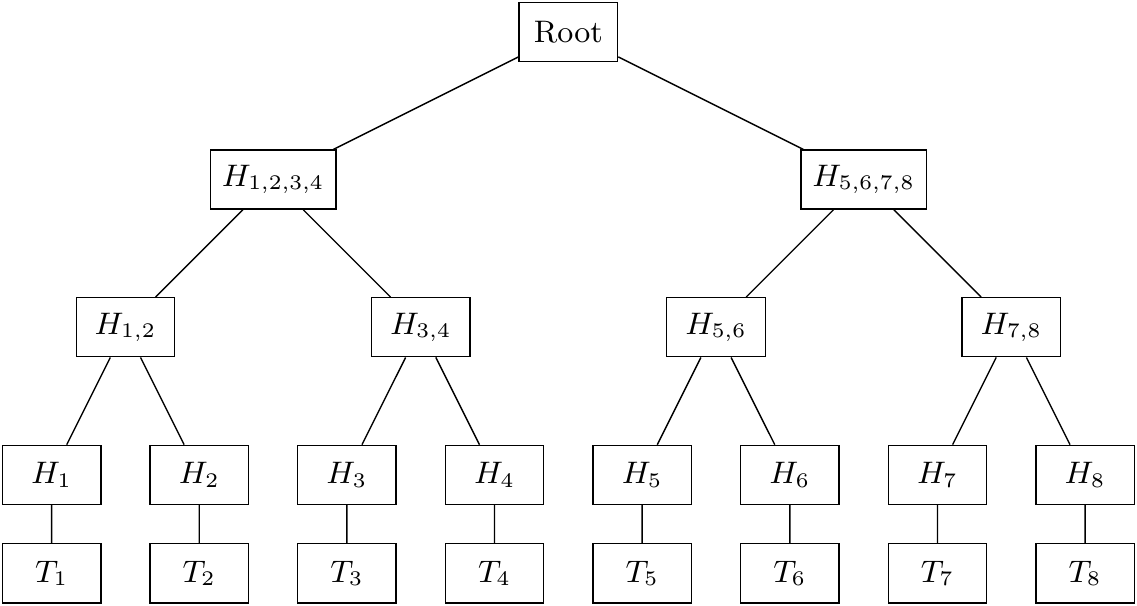}
\caption{Depiction of a Merkle tree. The leaf nodes, i.e. the elements of a Merkle tree, are written as '$T_i$. The non-leaf nodes are written as $H_j$.}
\label{fig:MT}
\end{figure}

A Merkle tree is a binary tree in which all leaf nodes (i.e. the Merkle tree's elements) are associated with a cryptographic hash, and all none-leaf nodes are associated with a cryptographic hash, that is formed from the hashes of its child nodes (as shown in figure \ref{fig:MT}).  

\begin{figure}[h]
\centering\includegraphics[width=0.7\linewidth]{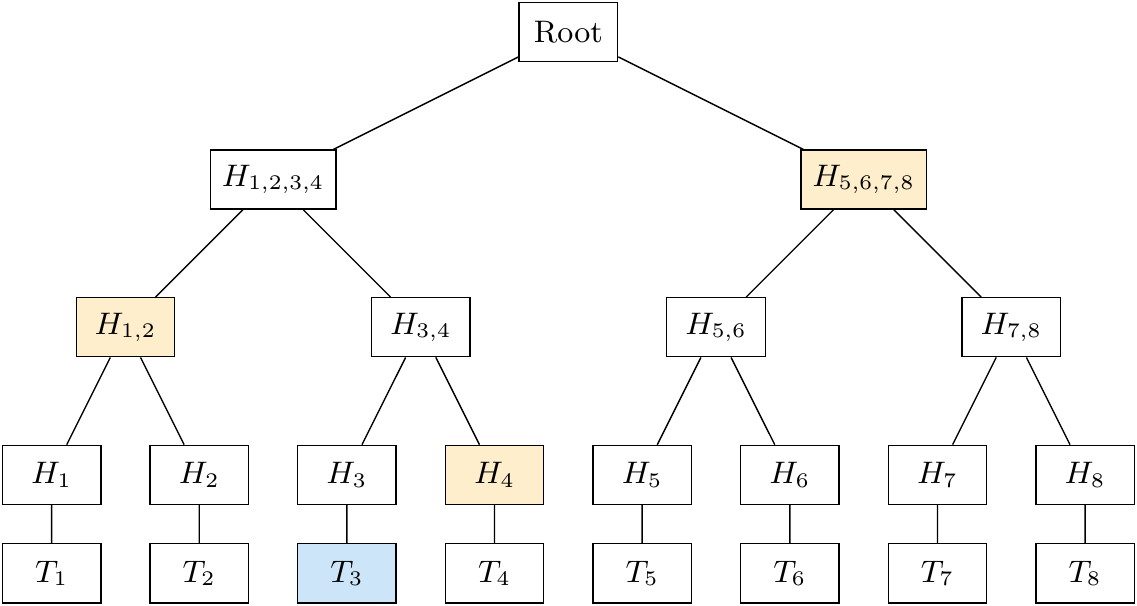}
\caption{Depiction of a Merkle tree, with the Merkle proof (shown in orange) for a given element ($T_3$).}
\label{fig:MT1}
\end{figure}

This simple structure allows for bandwidth-efficient and secure verification of the presence of elements. To verify that an element is present in the Merkle tree, one simply has to provide a series of hashes, that when hashed with the element hash, recreate the hash of the Merkle root (as shown in figure \ref{fig:MT1}). This series of hashes is also known as a Merkle proof. It is assumed that the recipient of the Merkle proof already has a copy of the Merkle root.

Merkle trees, like Bloom filters, have found a wide array of applications. The Merkle tree is used as a fundamental building block in blockchain systems \cite{NakamotoBitcoin:System}, in time synchronisation for batch signing requests \cite{RoughtimeGoogle}, in peer-to-peer key-value stores to check for the integrity of files \cite{BenetIPFS-Content3}, and more.

\subsection{Sparse Merkle Multiproofs}
\begin{figure}[h]
\centering\includegraphics[width=0.7\linewidth]{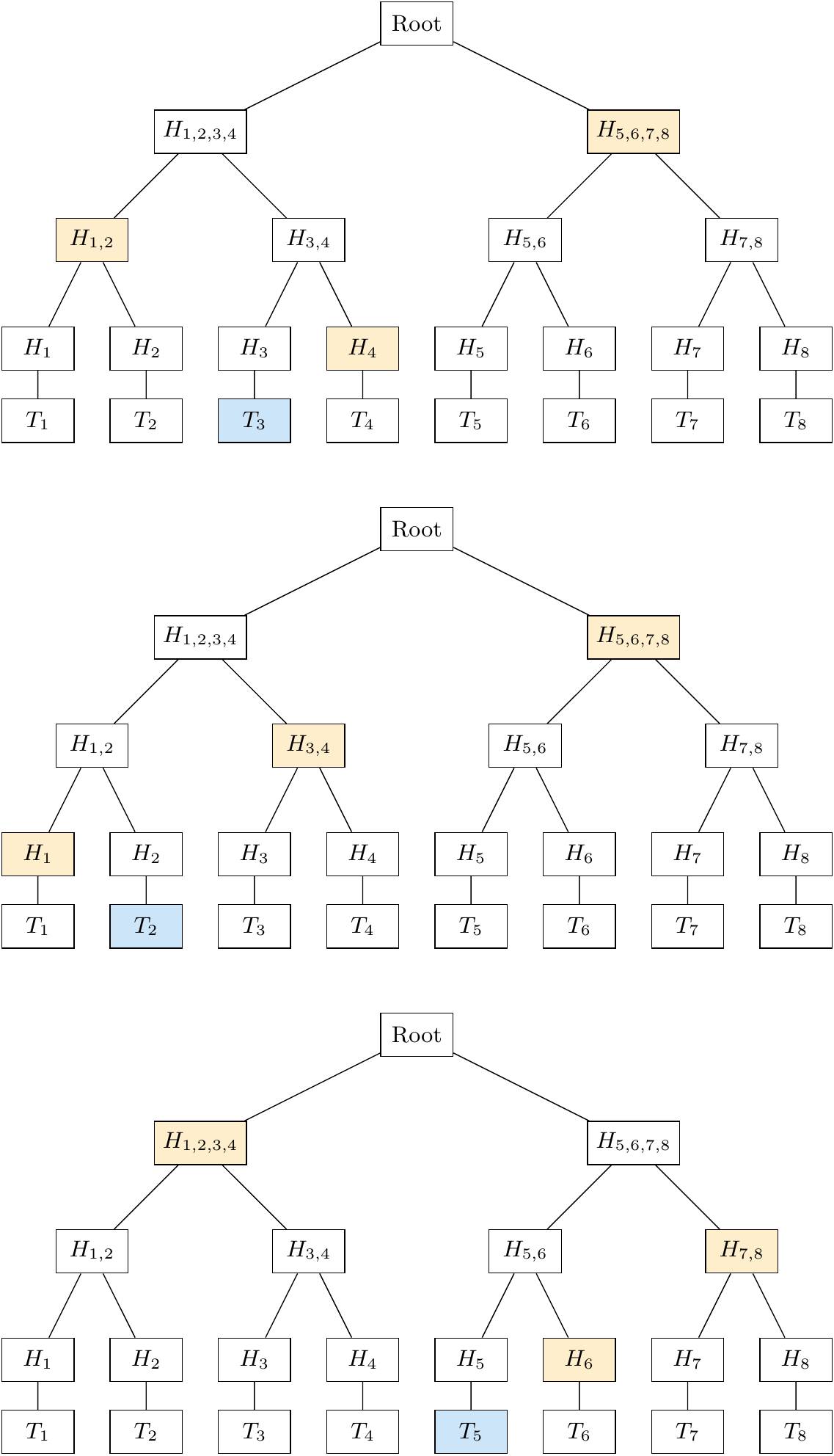}
\caption{Three Merkle proofs for three different elements.}
\label{fig:Merkle_three}
\end{figure}

A sparse Merkle multiproof (not to be confused with sparse Merkle trees) is simply a more efficient Merkle proof, for when it is necessary to prove the presence of a set of elements that are in the same Merkle tree \cite{UnderstandingTechnology}. As shown in figure \ref{fig:Merkle_three}, someone could use three Merkle proofs for the three elements highlighted in blue. In this example, a node would need nine hashes in total to verify the presence of the three elements. 

\begin{figure}[h]
\centering\includegraphics[width=0.7\linewidth]{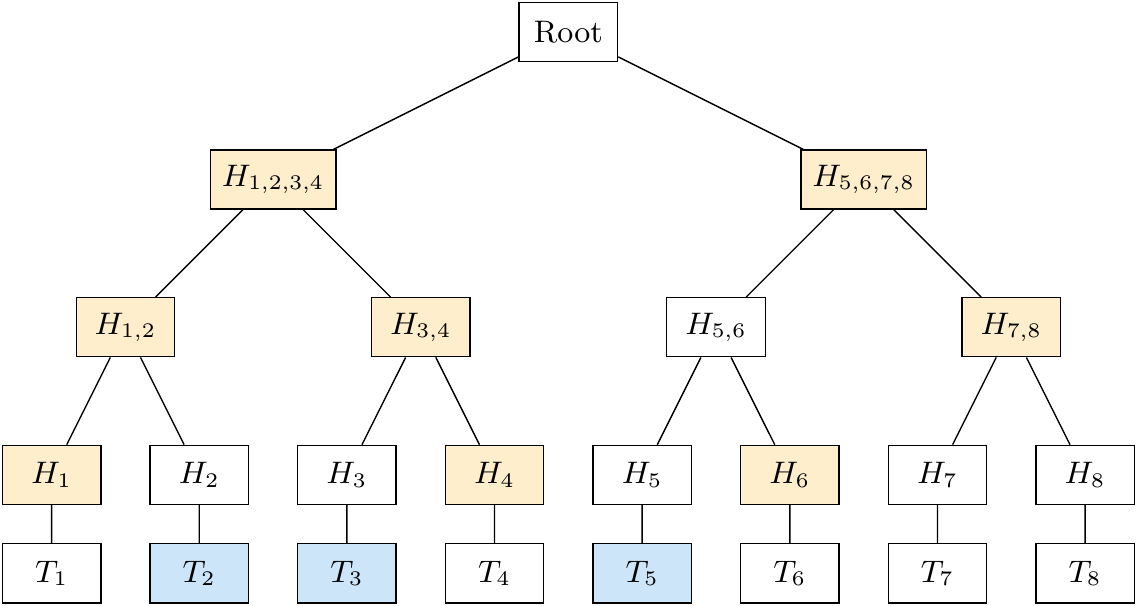}
\caption{Three overlapped Merkle proofs.}
\label{fig:Merkle_overlapped}
\end{figure}

\begin{figure}[h]
\centering\includegraphics[width=0.7\linewidth]{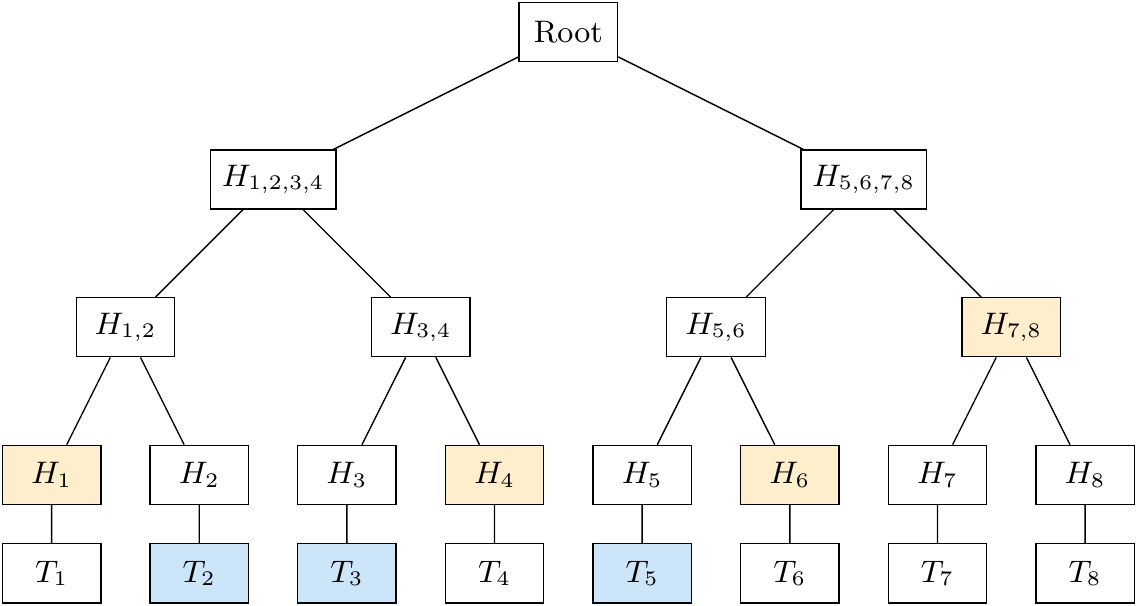}
\caption{An illustration of a Merkle multiproof.}
\label{fig:Merkle_multiproof}
\end{figure}

By using a sparse Merkle multiproof however, we can drop the number of hashes needed to verify the presence of a set of elements significantly. By overlapping the three trees from figure \ref{fig:Merkle_three} (as shown in figure \ref{fig:Merkle_overlapped}), we can see that most of the hashes can in fact be recreated by previous hashes. Instead of using three separate Merkle proofs that consist of nine hashes in total, one can prove the presence of the three elements with only four hashes (as shown in figure \ref{fig:Merkle_multiproof}. This simple trick is also known as a sparse Merkle multiproof.

\section{The  Bloom Tree}
The Bloom tree is a probabilistic data structure that combines the idea of Bloom filters with that of Merkle trees. In the standard Bloom filter, we  are interested to verify the presence, or absence of elements in a set. In the case of the Bloom tree, we are interested to verify and transmit the presence, or absence of an element in a secure and bandwidth efficient way to another party. Instead of sending the whole Bloom filter to check for the presence, or absence of an element, the Bloom tree achieves efficient verification by using a compact Merkle multiproof.

\begin{figure}[h]
\centering\includegraphics[width=1.\linewidth]{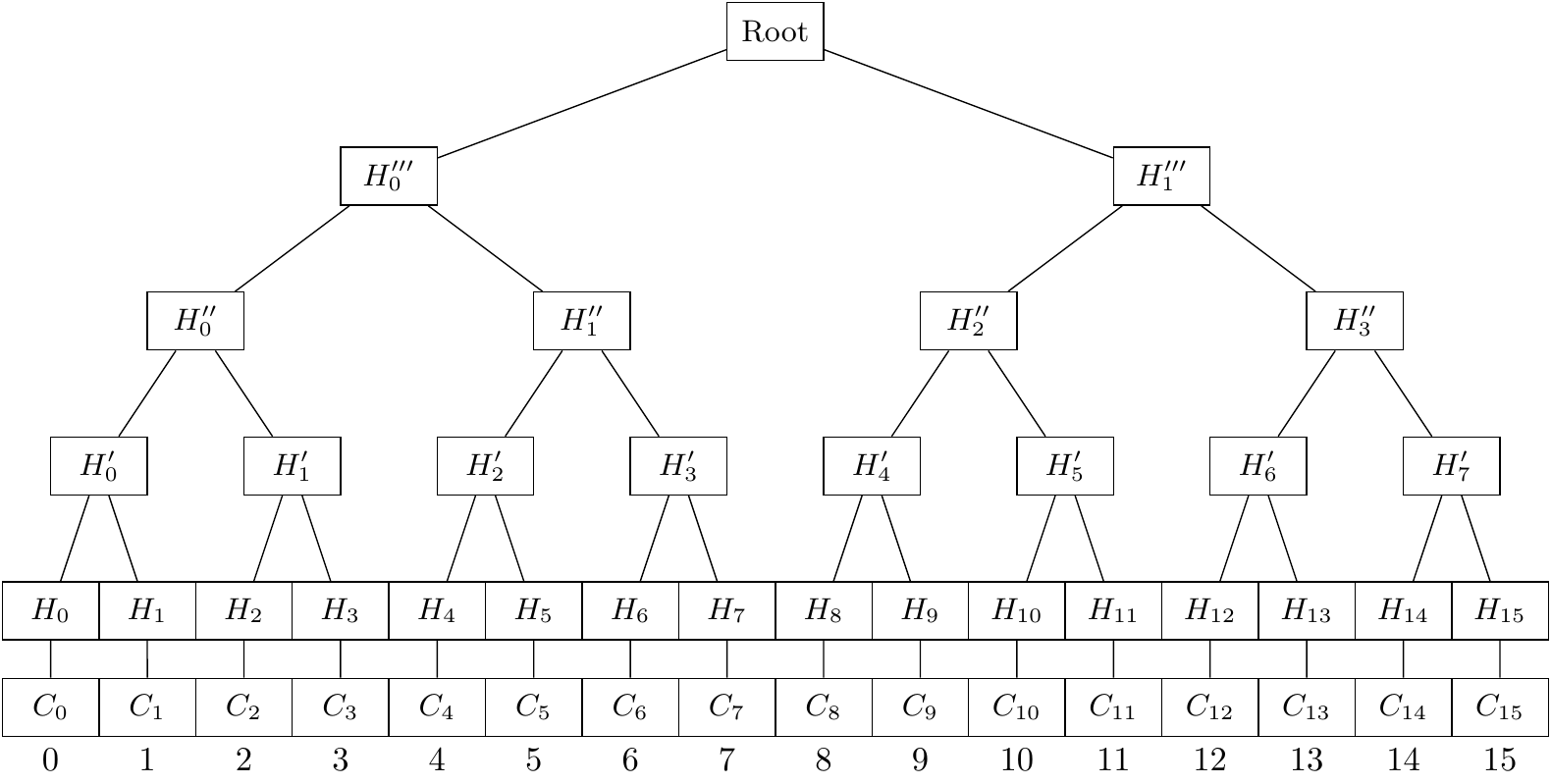}
\caption{The Bloom tree. Each cell $C_i$ represents a chunk of a Bloom filter.}
\label{fig:bt_1}
\end{figure}

The way a  Bloom tree is computed is straightforward:
\begin{enumerate}
    \item Divide the Bloom filter into chunks of bytes (8 bytes, 32 bytes, 64 bytes, or any other convenient size).
    \item Hash each chunk with its given chunk index.
    \item Compute a Merkle tree on top of the hashed chunks, as shown in figure \ref{fig:bt_1}.
\end{enumerate}{}

To prove if an element is present, or absent in the Bloom tree, we first hash a given element $k$ times. Let's say that we have a Bloom tree with a chunk size of $32$ bytes, and the $k$ hashes for a specific element are $\{800, 1602, 3650\}$. By knowing an element's $k$ indices and the Bloom tree's chunk size, we can deduce which chunks of the Bloom filter are needed to prove a given element inclusion, or non-inclusion. In our example, it would be chunks $C_3$, $C_6$, and $C_{14}$. We can also deduce which indices inside each chunk must be checked. In our case, to check index $800$, we would check index $31$ inside chunk $C_3$ (as one chunk contains $256$ bits in this example). For index $1602$ we would check index $65$ inside chunk $C_6$, and for index $3650$ we would check index $65$ inside chunk $C_{14}$.

\begin{figure}[h]
\centering\includegraphics[width=1.\linewidth]{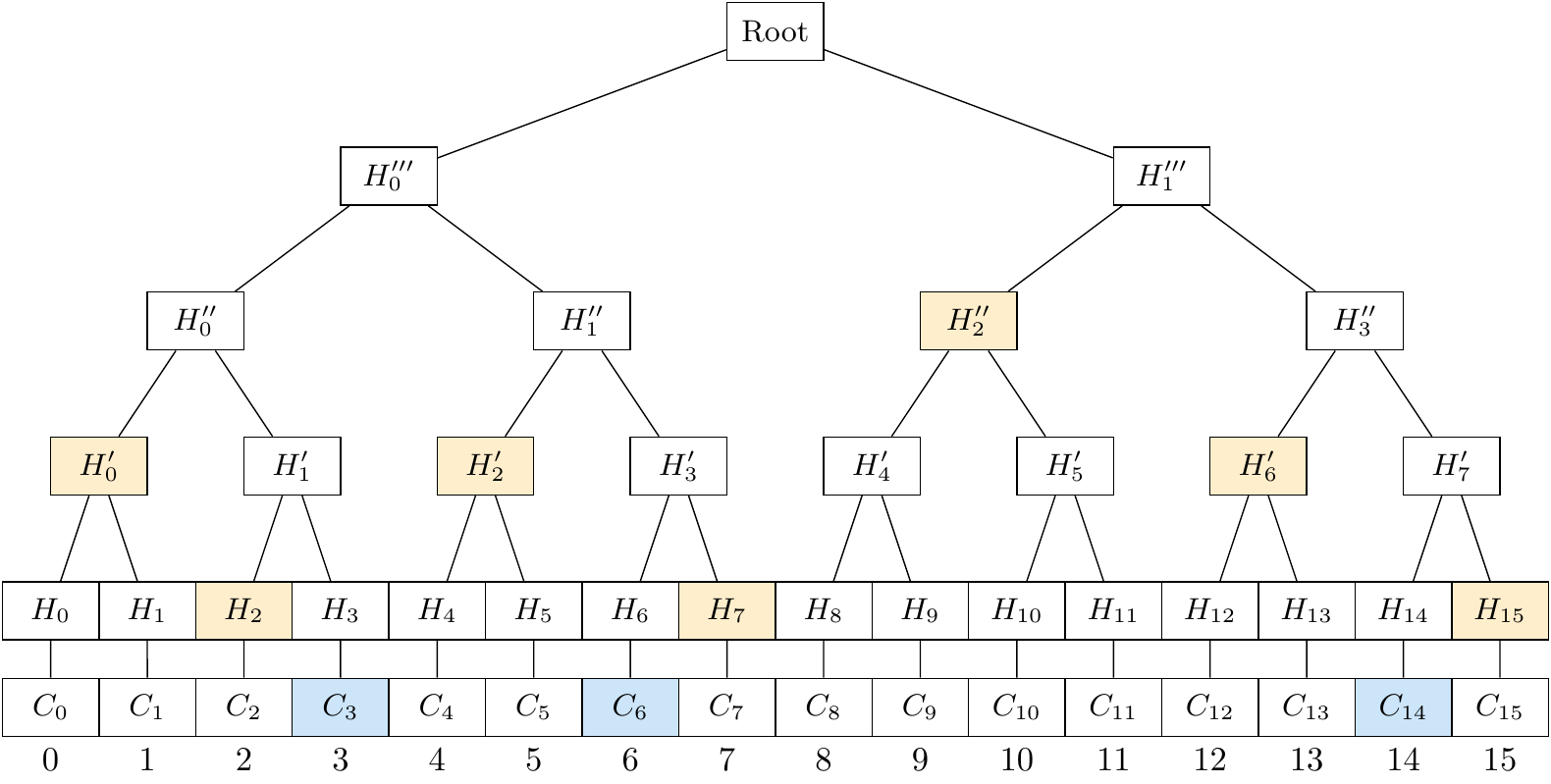}
\caption{The Bloom tree with a Merkle multiproof (shown in orange) for a certain element (shown in blue).}
\label{fig:bt_2}
\end{figure}

\begin{figure}[h]
\centering\includegraphics[width=1.\linewidth]{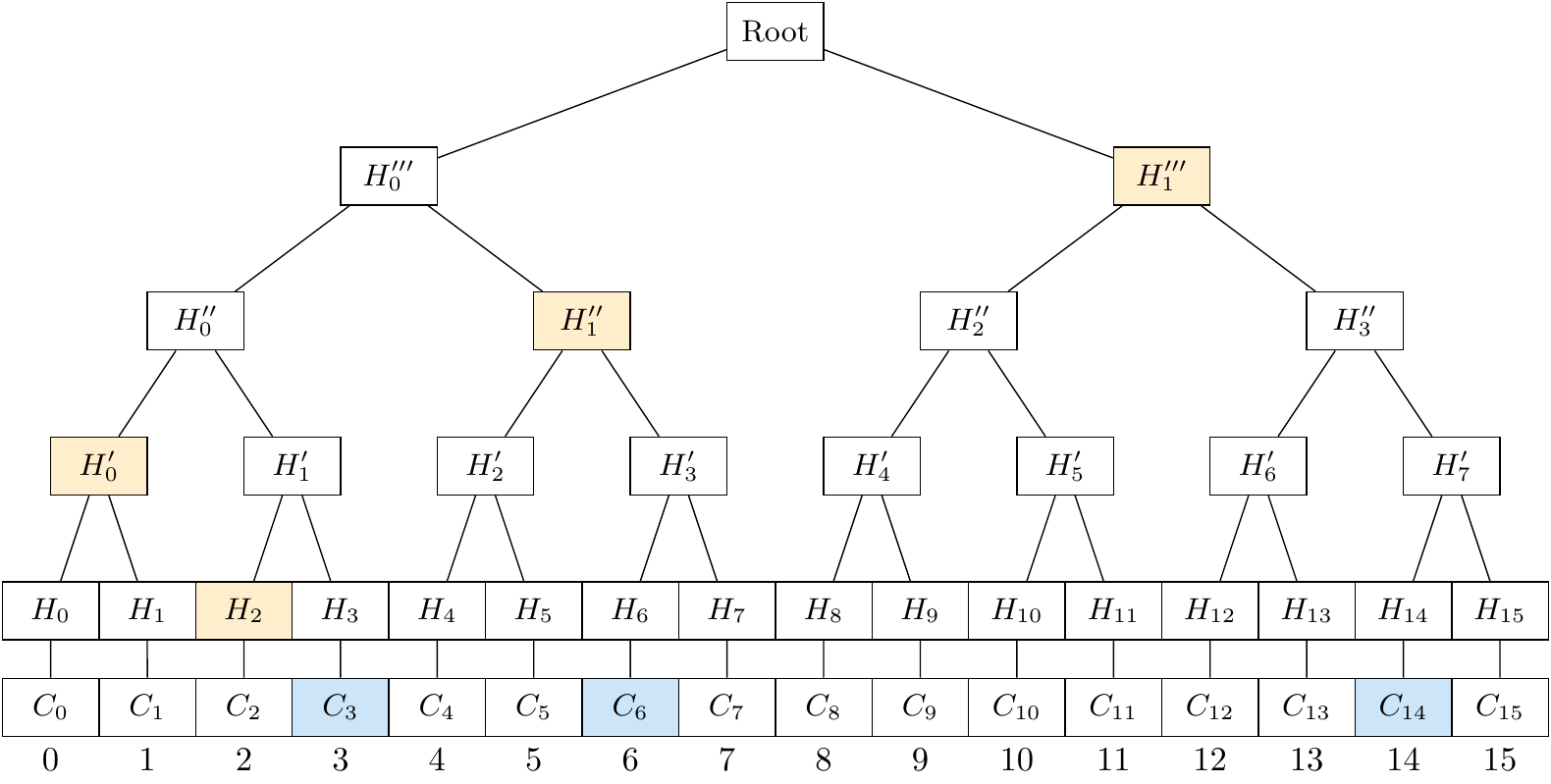}
\caption{The Bloom tree with a standard Merkle proof (shown in orange) for a certain element (shown in blue). In this particular case $C_3$ contained a zero at a given index. Instead of sending a Merkle multiproof, we only send a standard Merkle proof, as well as $C_3$ to prove the absence of the element.}
\label{fig:bt_3}
\end{figure}

If all of the values at the given indices are one, we provide a presence proof. If even only one of the values at the given indices is zero, we provide an absence proof. For a presence proof, we simply provide the Merkle multiproof for the given chunks, as shown in figure \ref{fig:bt_2}. Since each chunk gets hashed with its index, we know that the provided proof cannot be for another chunk (i.e. lying by providing a valid proof for another chunk is not possible). For an absence proof, we can simply provide a standard Merkle proof for one of the chunks in which the index gets mapped to a zero value, as shown in figure \ref{fig:bt_3}.

It is important to note that only false positives can occur in Bloom filters, false negatives are not possible. In other words a presence proof does not mean that an element was truly inserted into the Bloom filter, it means that it \textit{might} have been inserted. An absence proof on the other hand means that an element definitely has not been inserted into the Bloom filter.

\begin{figure*}
\centering\includegraphics[width=0.85\linewidth]{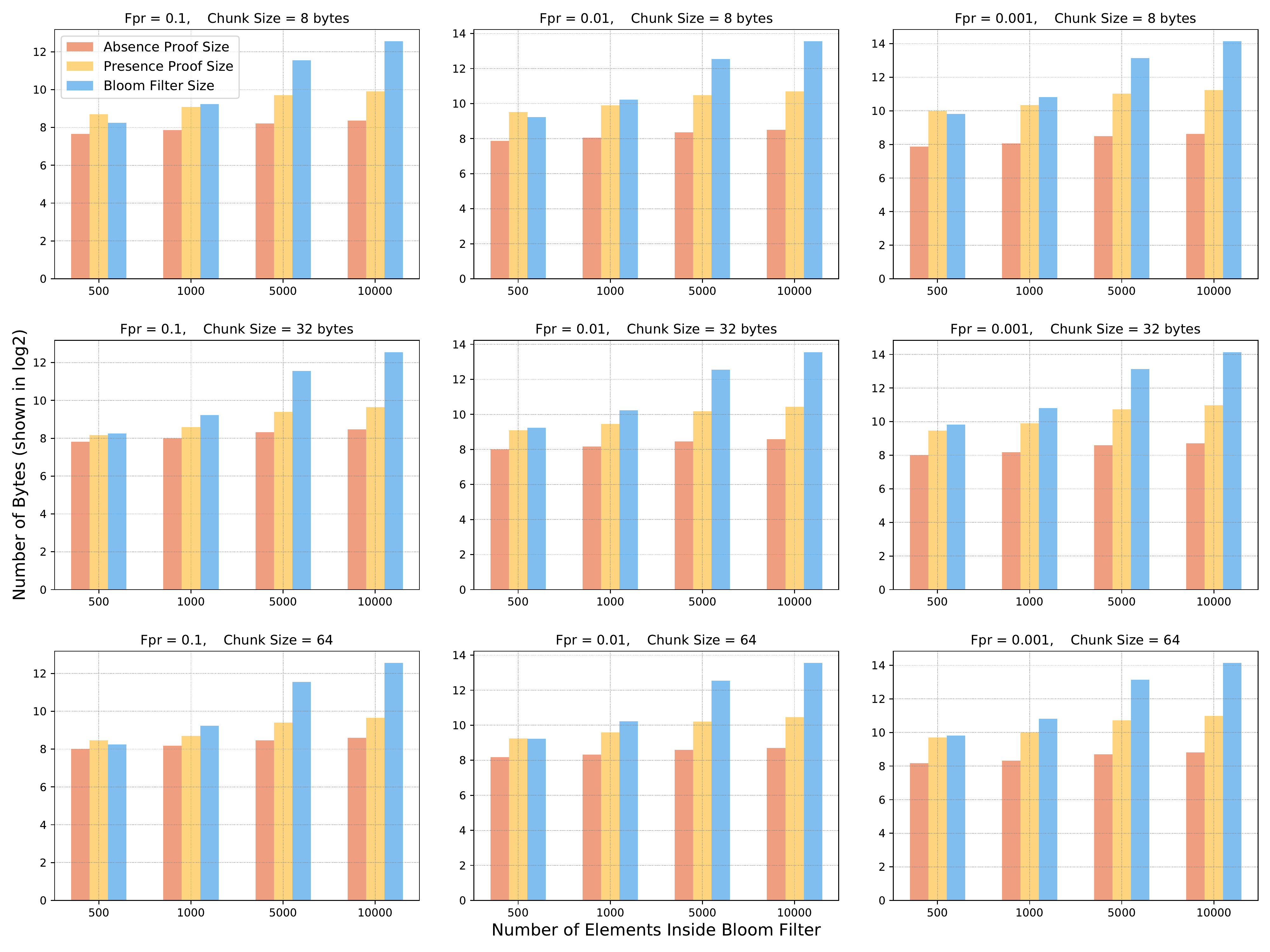}
\caption{Results of the experiments. Log2 scale was used for better visualization. A detailed explanation of the experiment setup is provided in subsection \ref{sub:exp_design}.}
\label{fig:plots}
\end{figure*}

\section{Experiments}
\subsection{Experimental Design}
\label{sub:exp_design}
We are interested to observe how the size of Bloom tree proofs varies, depending on the chosen Bloom filter parameters, as well as compare the proof sizes with their corresponding Bloom filters.
Nine grouped bar charts were computed, and organized as a grid (as shown in figure \ref{fig:plots}). Every bar chart in the same row used the same chunk size, and every bar chart in the same column used the same false positive rate. Chunks of size $\{8, 32, 64\}$ (in bytes) were used, and false positive rates of $\{0.1,0.01, 0.001\}$. For every bar chart four groups were computed. Each group compared the size differences between absence proofs, presence proofs, and the actual used bloom filter (measured in bytes). The only difference between each group was the parameter $n$ (the number of elements used to populate e bloom filter). The chosen number of elements for each group were $\{500, 1000, 5000, 10000\}$. As the size of a multiproof can be different from element to element, we computed the median multiproof size for each element from a sample of elements, i.e. every presence proof in the bar charts is the median size of a multiproof for a given bloom filter.

\subsection{Results}

Figure \ref{fig:plots} shows the results of the experiment. It appears that varying chunk sizes do not lead to a significant difference in proof sizes. As expected, larger chunk sizes will lead to slightly larger absence proofs, and slightly smaller presence proofs. The number of necessary computations however decreases with increasing chunk size. Intuitively, the larger a chunk size is, the fewer leaves there will be in the Bloom tree, which leads to fewer non-leaf computations. We can also see that the absence proofs and presence proofs get proportionally smaller than the Bloom tree, the larger the Bloom filter is. This also makes intuitive sense, as an absence proof (which is technically just a standard merkle proof) requires only $log2(m)$ hashes.

\section{Conclusion}
The Bloom tree's attributes appear to be of interest in two scenarios: 
\begin{enumerate}
    \item When for whatever reason we have to store regularly bloom filters.
    \item When we want to provide a probabilistic presence proof, or a non-probabilistic absence proof for individual elements to another party in a secure and bandwidth efficient way, instead of sending whole bloom filters.
\end{enumerate}{}
We believe that the Bloom tree will have particularly a lot of applications in the peer-to-peer and blockchain space.

\section{Future Work}
The Bloom tree package used for the experiments (which can be found on the Bloom Lab's \href{https://github.com/labbloom/bloom-tree}{\textcolor{blue}{github}} page) uses a specific variant of the sparse Merkle multiproof, which we name ''compact Merkle multiproof'' \cite{Ramabaja2020CompactMultiproofs}. In a nutshell, the compact Merkle multiproof is identical to a sparse Merkle multiproof, but it incorporates an algorithm for proof construction and verification, that allows us to send the hashes of a Merkle multiproof without requiring any additional information, such as hash indices.

In future work, we are going to show how one can combine Bloom trees with distributed Bloom filters \cite{Ramabaja2019TheFilter} to create an ''interactive Boom proof''. We will show how the interactive Bloom proof can be used to build a new kind of blockchain architecture, that requires one magnitude less storage, while still allowing nodes to independently verify transaction validity.

\addtolength{\textheight}{-12cm}   



\bibliographystyle{IEEEtran}




\end{document}